\documentclass[a4paper]{article}

\usepackage{INTERSPEECH2022}
\usepackage{tabularx}
\usepackage{booktabs}
\usepackage{multirow}
\usepackage{cleveref}
\usepackage{placeins}
\usepackage{siunitx}
\usepackage{todonotes}
\usepackage{url}
\newcommand{\fb}[0]{FluencyBank }

\title{Detecting Dysfluencies in Stuttering Therapy Using wav2vec 2.0}
\name{Sebastian P.~Bayerl$^1$, Dominik Wagner$^1$, Elmar Nöth$^2$, Korbinian Riedhammer$^1$}
\address{
  $^1$Technische Hochschule Nürnberg Georg Simon Ohm, Germany \\
  $^2$Friedrich-Alexander-Universität Erlangen-Nürnberg}
\email{sebastian.bayerl@ieee.org} 

\begin{document}

\maketitle
\begin{abstract}
\textit{Stuttering} is a varied speech disorder that harms an individual's communication ability.
Persons who stutter (PWS) often use speech therapy to cope with their condition. 
Improving speech recognition systems for people with such non-typical speech or tracking the effectiveness of speech therapy would require systems that can detect dysfluencies while at the same time being able to detect speech techniques acquired in therapy.

This paper shows that fine-tuning wav2vec 2.0 \cite{baevski_wav2vec_2020} for the classification of stuttering on a sizeable English corpus containing stuttered speech, in conjunction with multi-task learning, boosts the effectiveness of the general-purpose wav2vec 2.0 features for detecting stuttering in speech; both within and across languages.
We evaluate our method on \fb, \cite{bernstein_ratner_fluency_2018} and the German therapy-centric Kassel State of Fluency (KSoF) \cite{bayerl_ksof_2022} dataset by training Support Vector Machine classifiers using features extracted from the fine-tuned models for six different stuttering-related events types: blocks, prolongations, sound repetitions, word repetitions, interjections, and -- specific to therapy -- speech modifications. 
Using embeddings from the fine-tuned models leads to relative classification performance gains up to 27\% w.r.t. F1-score.  
\\
\end{abstract}
\vspace{-2mm}
\section{Introduction}

Stuttering is a varied speech disorder that affects about 1\,\% of the population \cite{yairi_epidemiology_2013}.
It is characterized by an increased duration and occurrence of dysfluencies compared to persons that do not have a stutter. 
The core symptoms of stuttering are repetitions, prolongations of sounds, syllables or words, and blocks while speaking \cite{lickley_disfluency_2017}. 
These can be accompanied by diverse linguistic, physical, behavioral, and emotional symptoms. 
Stuttering is not a static condition, and symptoms as well as severity vary significantly between speakers and even within the same speaker. 
The communication situation, psychological factors, and the linguistic complexity of an utterance influence the appearance of stuttering symptoms \cite{ellis_handbook_2009}. 
The symptoms interfere with the ability to communicate effectively and, therefore, negatively affect the life of a person who stutters (PWS).
The variability of the condition makes it very hard to detect reliably, and resources for training detection systems can hardly capture all aspects of stuttering.  
Detecting stuttering has implications for developing inclusive automatic speech recognition systems or self-help applications amongst other things.

Previous work has explored traditional signal processing, pattern recognition methods, and recent work has focused on detecting stuttering from speech with deep learning (DL) based approaches. 
For instance, \cite{noeth_automatic_2000} used Hidden Markov Models to detect stuttering in read speech. 
\cite{lustyk_language-independent_2015} could show that certain speech measures, such as the average length of silence that were designed and evaluated on Czech stuttering data, were transferable to German stuttering recordings. 
\cite{bayerl_towards_2020} did show that stuttering is detectable from the average duration of phonemes inside labeled dysfluency events.
Esmaili et al. did use Support Vector Machines (SVM) with  perceptual linear predictive (PLP) features to detect prolongations in speech \cite{esmaili_automatic_2017}.
More recent work was done by \cite{kourkounakis_detecting_2020}, who formulated stuttering detection as several binary classification problems and solved it using Residual Neural Networks (ResNet) in conjunction with bidirectional long short-term memory networks (LSTM). 
Whereas \cite{sheikh_stutternet_2021} used time delay neural networks (TDNN) to detect different kinds of stuttering in the UCLASS dataset \cite{howell_university_2009}.
Lea et al. used LSTM networks with multi-task learning to detect stuttering from podcast data \cite{lea_sep-28k_2021}.

In this paper, we show the applicability of \textit{wav2vec 2.0} (W2V2) as a feature extractor for dysfluency detection.
The diversity of the six stuttering-related event types we classify in this paper shows the relevancy of W2V2 in the context of dysfluency detection.
To show this, we use a pretrained W2V2 feature extractor and fine-tune multiple configurations of the W2V2 model for each dysfluency type, using either a single training objective or multi-task learning on a sizable English dataset containing stuttered speech. 
Thus, providing evidence for the complimentary of gender information as an auxiliary objective for multi-task learning when fine-tuning W2V2 models for dysfluency detection.
We evaluate our method on the \fb and the German Kassel State of Fluency (KSoF) dataset, providing further evidence for the transferability of features learned on English stuttered speech to other languages.
To the best of our knowledge, we are the first to extensively use and fine-tune wav2vec 2.0 models to detect six different stuttering-related events. 
Providing a valuable resource that can help narrow down the best location to extract speech representations from W2V2. 


\section{Data}\label{sc:data}
\subsection{SEP-28k}\label{ss:sep}
SEP-28k is a corpus containing English stuttered speech from podcasts \cite{lea_sep-28k_2021}. 
It consists of about 28000 clips extracted from podcasts that three annotators labeled with five types of dysfluencies: blocks, prolongations, sound repetitions, word repetitions, and interjections.
Making it one of the largest publicly available resources containing labeled stuttered speech. 
The corpus comes with an additional 4144 clips that were extracted from the adults who stutter part of the \fb corpus \cite{bernstein_ratner_fluency_2018}.
Those clips were labeled using the same protocol.
The authors claim that SEP-28k solves the problem of data scarcity  w.r.t. annotated audio containing stuttering as models trained on SEP-28k and a portion of \fb significantly outperform models trained and evaluated solely on \fb \cite{lea_sep-28k_2021}. 

As the corpus comes with little meta-data, besides the podcast's name and the episode the clip was taken from, we researched the missing gender label for the speakers of each episode from public podcast meta-data available. 
The additional meta-data was needed for multi-task learning experiments. 
The gender for the \fb corpus could be taken from the metadata provided by the original authors of \fb \cite{bernstein_ratner_fluency_2018}. 

In our fine-tuning experiments, we use $\sim${25000} clips from the SEP-28k dataset for training and $\sim$3000 clips for validation.
The splits are podcasts exclusive.
The dataset is potentially dominated by one female speaker, being in up to $\sim$46\% of all clips in the dataset, as she is the host of two podcasts featured in the dataset (WomenWhoStutter, HeStutters) \cite{bayerl_InfluenceDatasetPartitioning_2022}. 

When aiming for objective evaluation results, it is detrimental to have a test set that contains data from speakers that are in the training set. 
Considering the speaker distribution is essential when selecting a development or test set, because it will influence the results and generalization capability of the trained models \cite{bayerl_InfluenceDatasetPartitioning_2022}.
Stuttering is highly individual, and a PWS often stumbles over the same words, which could lead the system to learn speaker-specific traits, behaviors, and words, rather than stuttering. 
Random splitting clips without knowledge of the speaker label, will lead to overoptimistic results, especially if datasets are dominated by few speakers.  
Only in applying these best practices and unifying the evaluation of stuttering detection tasks, we will be able to assess the actual capabilities of stuttering detection systems \cite{bayerl_InfluenceDatasetPartitioning_2022}.

We use the \fb clips labeled in SEP-28k for the evaluation of the fine-tuned embeddings in SVM training. 
The data split we use consists of clips from six individuals ($\sim$800 clips) for testing and the remaining clips for training SVM systems with fivefold, speaker exclusive cross-validation on the remaining 26 individuals ($\sim$3300 clips).
Comparing results to other studies is difficult, as there is no fixed split for the new labels for the \fb data. 
The huge deviations up to $\pm$0.46 w.r.t. F1-score, in the results we got during the cross-validation experiments, lets us conclude that cherry-picking splits for good results is something to be mindful about.  
Therefore we made our speaker exclusive split publicly available.\footnote{\protect \url{https://tinyurl.com/24vm6dec}}
\subsection{Kassel State of Fluency}
The Kassel State of Fluency (KSoF) dataset is a resource containing recordings from 37 PWS in German.
The dataset contains 5597 3-sec long clips extracted from stuttering-therapy recordings.
The clips were labeled by three annotators with five types of dysfluencies: blocks, prolongations, sound repetitions, word repetitions, interjections, and an additional -- therapy-specific -- label, speech modifications \cite{bayerl_ksof_2022}.
The clips annotated as containing modified speech display a speech technique known as fluency shaping, which PWSs learn in stuttering therapy to help them overcome their stutter.
The labels and the data format are compatible with the datasets described in \cref{ss:sep}.
The label distribution w.r.t any of the dysfluencies is strongly imbalanced.
The dataset also comes with meta-data describing the speakers' gender, the recording device used, and the communication situation. 

In our classification experiments, we use the combined data from the train and development set containing 31 speakers (4344 clips) in five-fold cross validation and report results achieved on the test set comprised of recordings of eight speakers (1253 clips).  

\section{Methods}
\subsection{wav2vec 2.0}\label{ss:w2v2}

A common assumption is that neural networks need large amounts of labeled training data.
Arguably good generalization and model performance is not possible in tasks that lack large amounts of labeled training data. 
In recent years transformer-based architectures have achieved state-of-the-art performance in several fields. 
Those models are pretrained on large amounts of unlabeled training data.
First breakthrough results were achieved on natural language processing tasks. \cite{devlin_bert_2019}.  
Those models can be used as feature encoders, with or without adaptation. 
The extracted features can be utilized in downstream tasks, such as classification, as the models learned to extract many aspects of the underlying data. 

\begin{figure}[!htb]
    \centering
    \includegraphics[width=1.0\linewidth]{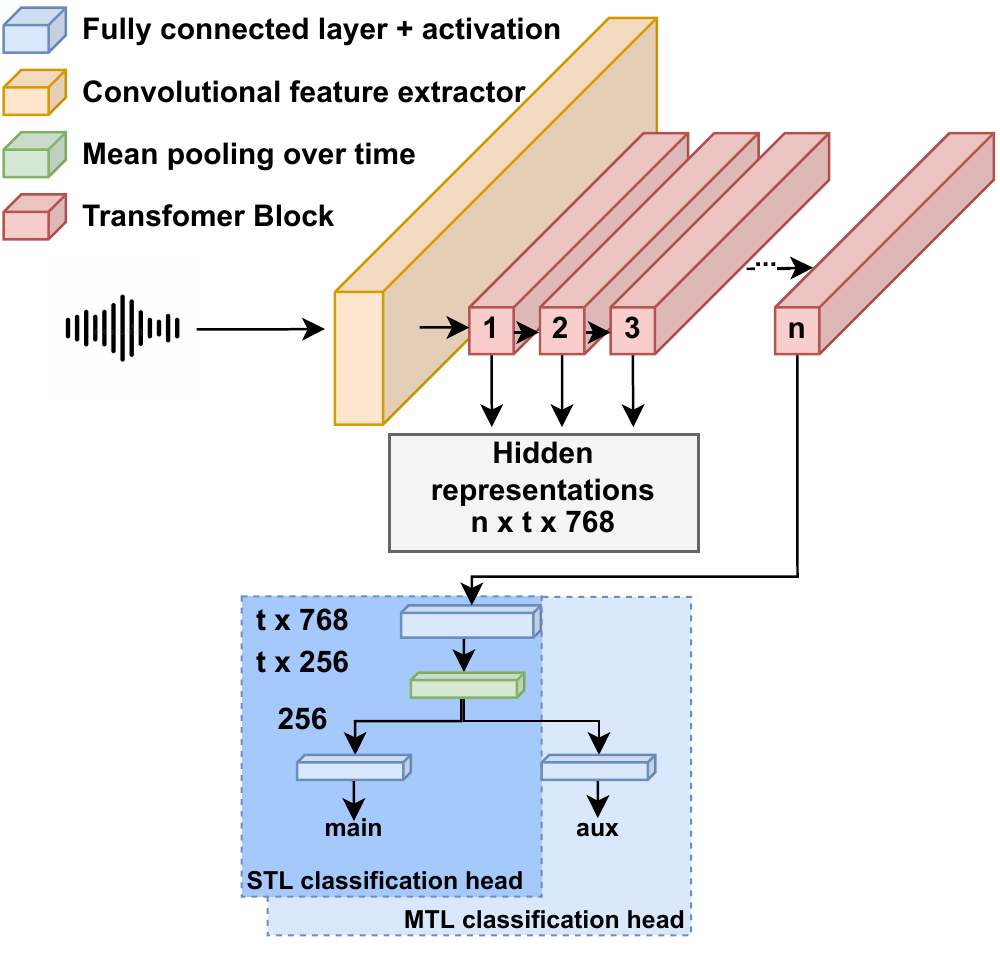}
  \vspace{-3mm}
    \caption{Schematic wav2vec 2.0 architecture overview with a mean-pooling-based single- or multi-task (STL/ MTL) classification head added after the $n_{th}$ transformer block.}
  \label{fig:architecture}
  \vspace{-4mm}
\end{figure}

The wav2Vec 2.0 (W2V2) approach learns a set of speech units from large amounts of data. 
The transformer model takes raw wave audio data as a direct input. 
The first part of the model consists of a convolutional neural network (CNN) encoder, followed by a contextualized transformer network and a quantization module.
The CNN produces latent representations directly from the wave-form inputs that are then discretized by the quantization module.
The architecture makes heavy use of self-attention and the transformer blocks were modeled to focus on the ``most important'' parts to represent speech in audio. 
The W2V2 feature encoder yields contextualized audio representations that have information encoded about a vector's relationship to others \cite{baevski_wav2vec_2020}.

W2V2 features, from pretrained models, work on several speech tasks, such as phoneme recognition, speech emotion recognition, and mispronunciation detection. \cite{baevski_wav2vec_2020,pepino_emotion_2021,xu_explore_2021}. 
Dysfluencies interrupt the normal flow of speech. 
Therefore, we expect that relationships to other parts of speech will be disturbed as a consequence, and embeddings extracted from dysfluent speech will differ from embeddings extracted from fluent speech.
We, therefore, hypothesize that these features are suitable for dysfluency detection.
Each model layer yields a different representation that might be more or less suitable for a task than a later or previous layer \cite{baevski_unsupervised_2021}. 
Detecting dysfluencies in stuttering therapy is a task that consists of detecting six different event types that vary significantly in their characteristics.
We, therefore, expected that we would need features from different layers for each event type.   

\vspace{-2mm}
\subsection{Transfer learning}\label{ss:transferlearning}
\vspace{-1mm}
Instead of using W2V2 only as a pretrained feature extractor, as described in \cref{ss:w2v2}, for dysfluency detection in stuttering therapy, we want to leverage the large quantity of labeled stuttering data from the SEP-28k dataset and learn improved representations for stuttering detection.
In our experiments, we used the Transformer implementation of W2V2 \cite{wolf_transformers_2020}.
Their implementation of a sequence classification head on top of the W2V2 base model uses a projection layer, reducing  dimensionality ahead of applying mean pooling over time.
The mean-pooled output is then fed into a classification layer that maps the outputs to the classification targets, as depicted in \Cref{fig:architecture}. 

The W2V2 base model has twelve stacked transformer blocks, after each of which the model yields hidden speech representations.
As each layer has different properties, we decided not just to add a classification head after the last transformer block.
Instead, we trained twelve models per dysfluency type.
Starting from the base model, we removed $12 - n$ transformer blocks and added a classification head after the $n_{th}$ transformer block, with $n \in \{1, 2, ... 12\}$.
In all our fine-tuning experiments, we froze the weight computation for the convolutional feature extractor that is located at the beginning of the W2V2 models. 
A schematic overview of the architecture employed can be found in \Cref{fig:architecture}.

\vspace{-2mm}
\subsection{Multi-task learning}
\vspace{-1mm}
An issue with training stuttering detection systems is overfitting and a lack of generalization \cite{sheikh_machine_2021}. 
Systems trained on podcasts data might perform poorly on therapy data and vice versa.
In low-resource scenarios like dysfluency detection in stuttering therapy, overfitting is particularly an issue. 
Multi-task learning (MTL) is a regularization method that helps to avoid overfitting.
Learning an additional, so called auxiliary task, that is different but related to the main task forces the model to generalize.  
The model can profit from cross-task regularization and auxiliary information. 
MTL has successfully been used in multiple speech tasks, such as speech recognition \cite{pironkov_speaker-aware_2016, ravanelli_multi-task_2020} or speech emotion recognition (SER) \cite{cai_speech_2021}.

Sheikh et al. suggested that MTL could be helpful to improve the generalization ability of stuttering detection systems and propose speaker- or gender recognition as auxiliary tasks \cite{sheikh_machine_2021}.
Lea et al. have already successfully employed MTL using two output branches in their networks. 
One branch features a fluent/dysfluent decision with a single cross-entropy term with focal loss; the other per-dysfluency branch used a concordance correlation coefficient (CCC) loss \cite{lea_sep-28k_2021}.
As about four times as many men than women are affected by stuttering \cite{yairi_genetics_1996}, we chose gender as our auxiliary task. 
We hypothesize that the training can benefit from the additional gender information while at the same time profiting from the regularizing effect. 

To implement this, we modified the classification head described in \cref{ss:transferlearning} by adding a second output layer of the same dimensionality, as can be seen in \Cref{fig:architecture}.
This output branch also takes the output of the pooling layer as its' inputs.  
We compute a weighted cross-entropy loss (CEL) for the auxiliary task.
The combined loss is calculated by \cref{eq:loss}, where $L_{\text{main}}$ is also a weighted CEL.
\begin{equation}\label{eq:loss}
    \mathbf{L}_{\text{MTL}} = w_{\text{main}} * L_{\text{main}} + w_{\text{aux}} * L_{\text{aux}}
\end{equation}
We experimentally found that weighing the auxiliary and the main task with $w_{\text{aux}} = 0.1$ and $w_{\text{main}}=0.9$ reduced the time it took the training to converge and minimized the overall value of the combined loss term $\mathbf{L}_{\text{MTL}}$.

\vspace{-2mm}
\section{Experiments}
All experiments described in this section follow the same generic approach. 
For each of the experiments, we extract features from a W2V2 model.
The W2V2 model employed were based on the W2V2 base model, which has 12 transformer blocks on top of the convolutional feature extractor \cite{baevski_wav2vec_2020}. 
Our models vary in three ways: the data they were trained on, the number of transformer blocks the model had during fine-tuning, and the kind of classification head used, either for single task learning (STL) or MTL. 

All experiments are formulated as binary classification tasks of one specific dysfluency against all other samples, containing speech of the other five stuttering event types and fluent speech. 
We report results using the F1-score for each of our experiments in \Cref{tab:results_high}.
We extracted a 768-dimensional speech representation for roughly every 20ms of raw audio, yielding 149 embeddings for each 3s audio clip. 
We computed the mean of all vectors per 3-sec clip from the extracted features, similar to the mean-pooling operation in the W2V2 classification head.  
These embeddings were then used as inputs for training SVM classifiers with radial basis function (RBF) kernels.
We used SVMs on the extracted W2V2 features to quickly explore the suitability of fine-tuned embeddings in different settings. 
SVMs can learn from only a few samples \cite{boser_training_1992}, as opposed to neural networks, and are therefore suitable for learning from small datasets, such as \fb and KSoF.

The optimal hyperparameters for the classifiers were determined using grid search in fivefold cross-validation on the respective training sets described in \cref{sc:data}.
We performed a principal component analysis (PCA) on the W2V2 embeddings prior to SVM training as part of the grid search. 
The number of principal components was chosen from $N_{\text{pca}} \in \{ 32, 64, 128, 256, 512\}$. 
The kernel parameter $\gamma$ was selected from the set $\gamma \in \{10^{-k} \mid k = -5, \ldots, -1 \} \subset \mathbb{R}_{>0}$, and
the penalty parameter of the error term $C$ was selected from $C \in \{ 1, 10, 100, 1000 \} \subset \mathbb{N}_{>0}$.

\vspace{-1mm}
\subsection{Baseline wav2vec 2.0}

The model we used for our initial experiments (W2V2-BASE) was pretrained in an unsupervised manner on 960 hours of unlabeled speech from the LibriSpeech corpus \cite{panayotov_librispeech_2015}.
It was subsequently fine-tuned for automatic speech recognition (ASR) on the transcripts of the same data. 
We use the weights published by \cite{baevski_wav2vec_2020}.
For our experiments we extracted hidden representations from all twelve transformer blocks and trained models with features from each layer. 
The best results for those experiments w.r.t the F1-score achieved on the respective test sets can be found in \Cref{tab:results_high} (W2V2-BASE) and results for different layers on the KSoF dataset are contained in \Cref{fig:w2v2performance}.

\vspace{-1mm}
\subsection{Fine-tuning wav2vec 2.0 using single-task learning} \label{ss:ft_stl}
Prior to extracting features from the W2V2 models in these experiments, we fine-tuned models for each type of dysfluency. 
For fine-tuning on the SEP-28k data, we use a batch size of 256 and fine-tune our models for up to 10 epochs, with a patience of 3 epochs. 
We used the adamW optimizer \cite{loshchilov_decoupled_2019} with an initial learning rate of \num{3e-5} and a warm-up phase consisting of 10\% of global training steps. 

After fine-tuning the models, we extract features at the $n_{\text{th}}$ layer.
Meaning that features from a model fine-tuned with a classification head added after the  $n_{\text{th}}$ transformer block were extracted only after this layer, not from any of the other still present $12 - n$ transformer blocks.
For modified speech, we use a meta-label, "any," during fine-tuning of the models.
The label is a combined label, containing all clips that were labeled as either of the five dysfluency types in SEP-28k.
We hypothesized that this would help the model learn the differences between ``normal'' and non-typical speech patterns. 
Results for the classification experiments trained with the embeddings gained from those models are reported in \Cref{tab:results_high} under W2V2-STL.


\subsection{Fine-tuning wav2vec 2.0 using multi-task learning} 

The W2V2 models for these experiments were fine-tuned using the combined MTL loss from \eqref{eq:loss} instead of a single CEL term.
The training parameters used for fine-tuning and the extraction scheme were identical to the ones described in \cref{ss:ft_stl}.
Results for these classification experiments can be found in \Cref{tab:results_high} (W2V2-MTL).


\begin{table}[!ht]
\vspace{-2mm}
    \centering
    \caption{Results of binary classification of dysfluency events. We report the F1-score for each dysfluency class for \fb and KSoF. (\textbf{Mod} = Modified Speech, \textbf{Bl} = Block, \textbf{Int} = Interjection, \textbf{Pro} = Prolongation, \textbf{Snd} = Sound repetition, \textbf{Wd} = Word repetition)}
    \vspace{-2mm}
    \scalebox{0.95}{
\begin{tabular}{l|c|c|c|c|c|c}
\toprule
  \textbf{System} &  \textbf{Mod}  &  \textbf{Bl} &  \textbf{Int} &  \textbf{Pro} &  \textbf{Snd}  & \textbf{Wd} \\ 
    \midrule
 \multicolumn{7}{c}{\textbf{\fb}}   \\
    \midrule
  \textbf{W2V2-BASE}&  -  &  0.30  &  0.70  &  0.51  &  0.50  & 0.39  \\ 
  \textbf{W2V2-STL} &  - &  0.31 & 0.83 & 0.52       &   0.40  & 0.40 \\ 
  \textbf{W2V2-MTL} &  - &  \textbf{0.33} &  \textbf{0.84}  & \textbf{ 0.60} & \textbf{0.60 }& \textbf{0.43} \\ 
    \midrule
\multicolumn{7}{c}{\textbf{KSoF}}   \\
    \midrule
 \textbf{W2V2-BASE} & 0.68  &  0.51  &  0.66  &  0.49  &  0.37  & \textbf{0.22}  \\ 
 \textbf{W2V2-STL}  & 0.73  &  0.49 & 0.71 & 0.49  &  0.43  & 0.18 \\ 
 \textbf{W2V2-MTL}  & \textbf{0.76} & \textbf{0.54} & \textbf{ 0.74}  &  \textbf{0.53} & \textbf{0.47} & 0.19 \\ 
\bottomrule
\end{tabular}
}
    \label{tab:results_high}
\vspace{-4mm}
\end{table}

\section{Discussion}
\begin{figure}[!htb]
\vspace{-3mm}
    \centering
    \includegraphics[width=\linewidth]{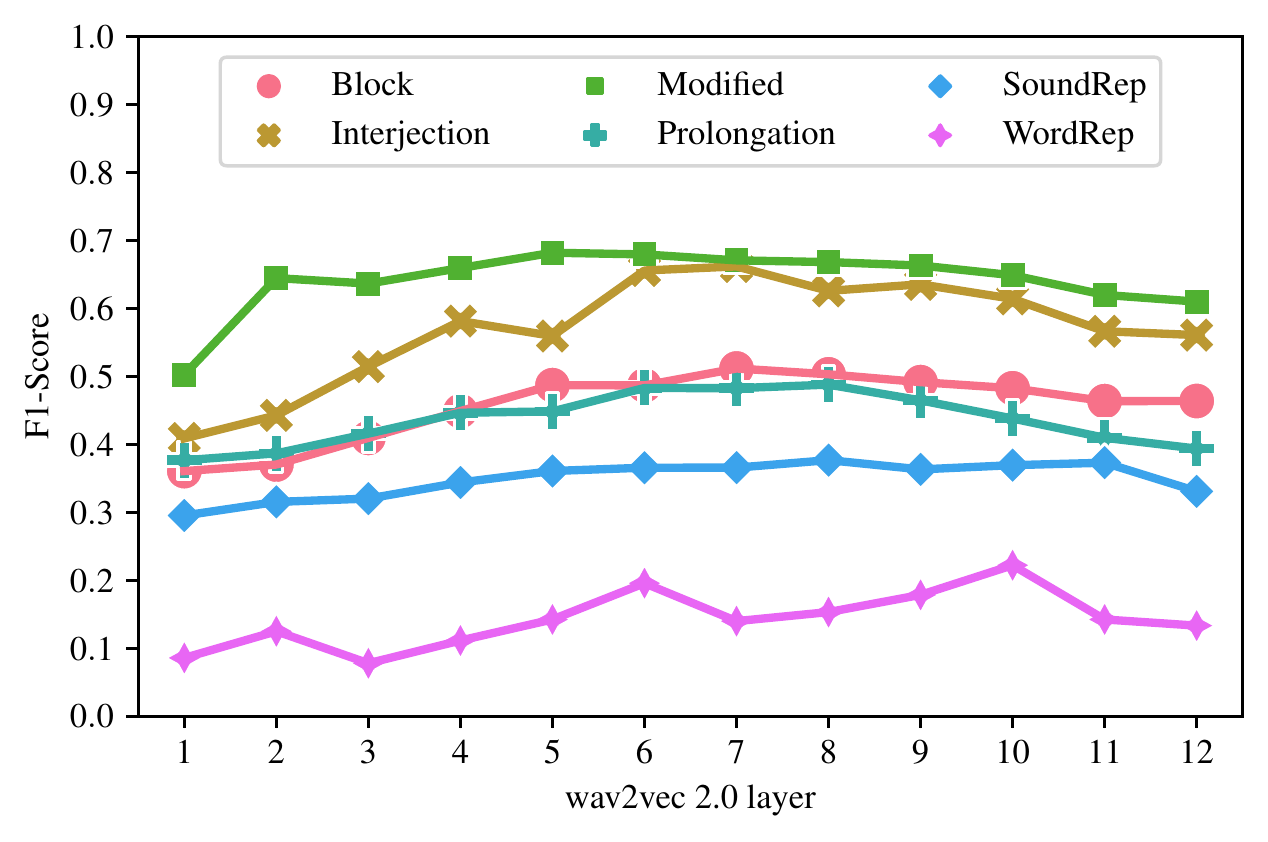}
    \caption{F1-score by wav2vec 2.0 (W2V2) extraction layer for each dysfluency type in KSoF computed with features extracted from the W2V2-BASE model. }
  \label{fig:w2v2performance}
  \vspace{-4mm}
\end{figure}

\Cref{fig:w2v2performance} shows the classification performance for systems trained on all W2V2 layers. 
Performance on all dysfluency types peaks for the features extracted between layers 5 and 10.
Features from early layers in the processing hierarchy, yield features that are less distinctive for dysfluency patterns, as dysfluency patterns are affecting speech directly, not low-level descriptors.
This is in agreement with results from \cite{baevski_unsupervised_2021}, where the middle layers instead of the top layers performed best in phoneme recognition tasks.

The results in \Cref{tab:results_high} show that classifiers for interjections, prolongations, and sound repetitions,  on both datasets profit substantially from the fine-tuned embeddings.
The classification performance for five of the six event-types profits from embeddings fine-tuned on stuttering in another language. 
The effect that the embeddings extracted from the W2V2-MTL models have w.r.t. the classification performance is positive across all experiments, but the word repetition experiment on KSoF, where fine-tuning, in general, seems to harm the already weak performance. 
Word repetitions need the largest acoustic context to be recognizable and repeated words are acoustically barely distinguishable from normal speech.
Averaging over all vectors extracted from the 3s clips removes much information, w.r.t. to the time-series character of speech, which probably makes classification harder.
The validity of the results for word repetitions on KSoF might also be limited, as only 3.8\% of the clips in KSoF were labeled as word repetitions \cite{bayerl_ksof_2022}. 

The classification performance w.r.t. the F1-score achieved on classifying blocks on the KSoF dataset is much better than on \fb. 
A possible explanation is that the PWS who currently undergo therapy are heavy stutterers who experience more severe blocks, making them easier to detect. 
A detailed spectral and prosodic analysis of this phenomenon is needed to shine some light on these differences. 



Interjections are a common functional dysfluency, found in PWS and people who do not have a stutter.  
The phonetic composition of those functional dysfluencies is similar between English and German \cite[p.17]{belz_phonetik_2021}, which probably leads to distinguishable features that transfer well between languages and therefore leads to good results on both datasets.
The classification of modified speech works consistently best for all features and event-types in the KSoF dataset. 
As this speech pattern is uniformly taught and acquired by practicing under the supervision of speech pathologists, it can be assumed that the pattern has less variance between speakers and is therefore easier to distinguish.
Interestingly, even though the fine-tuned W2V2 models never saw any modified speech, the classification performance on the extracted features still profited substantially from fine-tuning the W2V2 model on non-typical speech.

\vspace{-1mm}
\section{Conclusion}
We showed that pretrained and fine-tuned W2V2 models yield features suitable for dysfluency detection using SVMs. 
We showed that the classification performance on both datasets improves significantly when using features from systems fine-tuned on out-of-domain stuttering data, even across languages and scenarios.  
Multi-task learning has a regularizing effect on training W2V2 feature extractors and improves generalization. 
Overall classification performance on the \fb dataset appears better, which could be due to differences in language and or greater variety in recording situations and conditions in the KSoF dataset.

In future work, we plan to improve the feasibility of W2V2 features for all kinds of dysfluency events and extend our work to the multi-class scenario, not only  detecting, but differentiating multiple kinds of dysfluencies.
One way to achieve this will be to explore other feature aggregation strategies using an attention mechanism in conjunction with a classification token instead of mean pooling \cite{devlin_bert_2019,vaswani_attention_2017}.

\pagebreak
\newpage
\bibliographystyle{IEEEtran}

\bibliography{references,refs}

\end{document}